\documentstyle[12pt]{article}
\begin{document}

\noindent P.~N.~Lebedev Institute Preprint     \hfill
                                        FIAN/TD/ 20-92\\
I.~E.~Tamm Theory Department       \hfill
\begin{flushright}{Desember 1992}\end{flushright}
\begin{center}
\vspace{0.1in}{\Large\bf  SU(2)-EFFECTIVE ACTION WITH THE NONANALYTIC TERM}
\vspace{0.3in}\\
{\large  O.~K.~Kalashnikov}\\
\medskip  {\it Department of Theoretical Physics} \\ {\it  P.~N.~Lebedev
Physical Institute} \\ {\it Leninsky prospect, 53, 117 924, Moscow,
Russia}\footnote{E-mail address: tdparticle@glas.apc.org}\\

\end{center}

\vspace{1.5cm}

\centerline{\bf ABSTRACT}

\begin{quotation}

The nonanalytic $g^3$-term is calculated for SU(2)-effective action at
finite temperature and the status of a gauge fields condensation
is briefly discussed.
\end{quotation}

\newpage

The gauge models with embeded external fields are a very important object
of modern theoretical physics because their essence is rather realistic
and adequate to many physical phenomena. The gauge fields condensation
which can be simulated through a very simple external field is a typical
phenomenon for many unified gauge models and its properties are
intensively studied today by using mainly the effective action technique.
However two-loop effective action calculated with a nonzero external
field (at first in [1] and then for an arbitrary gauge parameter $\xi$
in [2] for SU(2)-group and in [3,4] for SU(3)-one) displays
that all its minima are physically equivalent although no evident
reasons exist to consider this degeneracy being proved. This fact makes
the phenomenon (where the gauge fields condensate arises spontaneously
[5]) unreliable and it is very important to establish the real
value of this scenario. Moreover the special attention should be paid
to a gauge-invariance of the results found [4,6], especially a situation
is not clear when the high-orders corrections being taken into account.
The search of these corrections (at least the $g^4$-order ones) is a
very actual task since only they can determine the status of this
phenomenon and a role of degeneracy in building the nontrivial vacuum.
Of course, the found degeneracy is partially determined by a symmetry
of the breaking operator which distroys the initial gauge group but
as concerns a trivial vacuum this degeneracy is false and results
from the unperfection of the calculational scheme
based on the lowest orders perturbative graphs.

The quantum SU(2)-Lagrangian in the background gauge has the standard form
$$
{\cal L}=-\frac{1}{4}(G_{\mu\nu}^a)^2-\frac{1}{2\xi}
[({\bar {\cal D}}_{\mu}V_{\mu})^a]^2 +
{\bar C}{\bar {\cal D}}_{\mu}{\cal D}_{\mu}C
\eqno{(1)}
$$
where the gauge fields $ V_{\mu}^a $ are decomposed in the quantum part
$ Q_{\mu}^a $ and the classical constant one $ \bar {A}_{\mu}^a $ (here
$ V_{\mu}^a=Q_{\mu}^a+\bar {A}_{\mu}^a $ ). The gauge fields strength
tensor is determined through the new covariant
derivative $ {\bar {\cal D}}_{\mu}^{ab}
=\partial_{\mu}{\delta}^{ab}+gf^{acb}\bar {A}_{\mu}^c $
and the derivative ${\cal D}_{\mu}^{ab}$ depends on $V_{\mu}^a$
in the usual way. The parameter $ \xi $ fixes the internal
gauge and the classical field has the form
$$
 \bar {A}_{\mu}^a=\delta_{\mu4}\delta^{a3}A^{ext}=
 \delta_{\mu4}\delta^{a3}(\frac{\pi T}{g} x)
\eqno{(2)}
$$
where $x$ is a new convenient variable. Here $T$ is temperature and
$g$ is the standard coupling constant.

The effective action for this model (including the two-loop graphs) has
been calculated by many authors [1,2] and has a rather simple form
\setcounter{equation}{2}
\begin{eqnarray}
&&W(x)/T^4=W^{(1)}(x)/T^4+W^{(2)}(x)/T^4,\nonumber\\
&&W^{(1)}(x)/T^4=\frac{2}{3}\pi^2[B_4(0)+2B_4(\frac{x}{2})],\nonumber\\
&&W^{(2)}(x)/T^4=\frac{g^2}{2}[B_2^2(\frac{x}{2})+2B_2(\frac{x}{2})B_2(0)]
+\frac{2}{3}g^2(1-\xi)B_3(\frac{x}{2})B_1(\frac{x}{2}),\nonumber\\
\end{eqnarray}
where $ B_n(z) $ are the modified Bernoulli polynomials
\begin{eqnarray}
&&B_1(z)=z-\epsilon(z)/2\,\,\,,\qquad
B_3(z)=z^3-3\epsilon(z)z^2/2+z/2,\nonumber \\
&&B_2(z)=z^2-|z|+1/6\,,\quad B_4(z)=z^4-2|z|^3+z^2-1/30 \nonumber\\
\end{eqnarray}
with $\epsilon(z)=z/|z|$. Here we should consider
that $\epsilon(0)=0$ as it results from the direct calculations to
make (3) be correct.

The action (3) has three extremum points
$$
\bar{x}=0\,,\qquad \bar{x}=1\,,\qquad \bar{x}=2,\\
\eqno{(5)}
$$
where two of them ($\bar{x}=0$ and $\bar{x}=2$ ) are minima of the
presented action. The effective action being put on these extremum points
is a gauge independent quantity [4] but, unfortunately, the thermodynamical
potential found within this approximation for the trivial vacuum
($ \bar{x}=0 $) and for the nontrivial one ($ \bar{x}=2 $) has the
same value
$$
\Omega/T^4=2{\pi^2}B_4(0)+\frac{3g^2}{2}B_2^2(0)=
-\frac{\pi^2}{15}+\frac{g^2}{24}
\eqno{(6)}
$$
This fact indicates that a degeneracy
(which is probably a signal of the real
one) takes place within this scenario and the multi-loop corrections
are very essential for clearing the situation. However the direct
calculation of a three-loop effective action (the $g^4$-order) is a
hopeless task and therefore a nonperturbative scheme
should be built to define the status of this phenomenon. Below the
simplest summation is used to calculate the leading nonanalytic term in
the nonperturbative expansion of $W(x)$ and we discuss its gauge
dependence. This term is the $g^3$-order and for many physical
phenomena plays a more essential role then the $g^4$ -terms to come.

It is well known (see e.g.[7,8]) that for any non-Abelian gauge theory
(despite of its more complicated structure) the leading nonanalytic term
can be reproduced through the standard formula
$$
\frac{\partial{W^{(cor)}(x)}}{\partial{g}}=\frac{1}{\beta g}
\sum_{k_4}\int \frac{d^3 k}{(2\pi)^3} \rm {Tr}[{\cal D}(k) {\Pi}(k)]
\eqno{(7)}
$$
where the polarization tensor $ \Pi (\bar{k},k_4) $ should be calculated
in the lowest order. For this calculation only $\Pi_{44}(|\bar{k}|\to 0,0)$
is used and the final result has the form
$$
\Delta W^{(cor)}=-\frac{\Pi_{44}^{3/2}(0)}{12\pi\beta}\rm {Tr}(I)
\eqno{(8)}
$$
Here I is the unit matrix in the adjoint representation of the chosen
gauge group (for SU(N) one has $\rm {Tr} (I)=N^2-1$).

Polarization tensor for the broken SU(2)-group (when $x\ne 0$) has two
components $\Pi^{||}(\bar{k},k_4)$ and $\Pi^{\perp}(\bar{k},k_4)$ which
are completely independent within the
$g^2$-approximation. Their calculations are standard and exploit the usual
temperature Green functions technique in the imaginary time space. To
simplify what follows all details used are omitted and below
the infrared limits of the $\Pi_{44}(\bar{k},k_4)$-components
are presented only by their leading terms.

The order $g^2$ infrared limit of $\Pi_{44}^{||}(\bar{k},k_4)$
has the simple form
$$
\Pi_{44}^{||}(|\bar{k}|\to 0,k_4=0)=4g^2T^2B_2(\frac{x}{2}),
\eqno{(9)}
$$
and for its calculating the standard prescription is used (here $k_4=0$
and then $|\bar{k}|\to 0$). It is very important to notice
that only the expression (9) is
generated by the effective action (3) through the usual formula
$$
m_{||}^2=\frac{g^2}{\pi^2T^2}\frac{1}{4}
[\partial^2 / \partial{(\frac{x}{2})^2}]W(\frac{x}{2})
\eqno{(10)}
$$
and there is a possibility to improve (9) up to the order $g^4$ terms
\setcounter{equation}{10}
\begin{eqnarray}
m_{||}^2&=&4g^2T^2B_2(\frac{x}{2})\nonumber\\
&+&\frac{g^4T^2}{\pi^2}\left\{B_1^2(\frac{x}{2})+\frac{1}{2}
[B_2(\frac{x}{2})+B_2(0)]+(1-\xi)[B_1^2(\frac{x}{2})+B_2(\frac{x}{2})
]\right\}\nonumber\\
\end{eqnarray}

The infrared limit of $\Pi_{44}^{\perp}(\bar{k},k_4)$ cannot
be found within (10) and it is calculated directly through
the Green functions technique. Moreover there are some peculiarities
which complicate a search of this limit when $x\ne 0$ since the initial
gauge symmetry is broken. In the transversal sector all gauge bosons
acquare a mass (a nonzero damping at the tree level)
and the infrared limit of
$\Pi_{44}^{\perp}(\bar{k},k_4)$ should be determined near a new mass shell
$\hat {k}_4=0$ (where $\hat{k}_4=k_4+{\pi}Tx$ ). The calculations
are standard and the order $g^2$ infrared limit of
$\Pi_{44}^{\perp}(\bar{k},k_4)$ has the form
$$
\Pi_{44}^{\perp}(|\bar{k}|\to0,\hat{k}_4=0)
=2g^2T^2 \left(B_2(\frac{x}{2})+B_2(0)\right)
\eqno{(12)}
$$
which is a gauge-invariant quantity for any $x\ne 0$. This is not the
case when all other possible infrared limits of
$\Pi_{44}^{\perp}(\bar{k},k_4)$ are studied and we consider that
only the expression (12) should be used within formula (8).

Now gathering all expressions found for the infrared limits
of $\Pi_{44}(\bar{k},k_4)$ and using formula (8) we obtain
the nonanalytic corrections as follows
$$
\Delta W^{(cor)}/T^4=-\frac{2g^3}{3\pi}\left\{B_2^{3/2}(\frac{x}{2})+
2[\frac{1}{2}(B_2(\frac{x}{2})+B_2(0))]^{3/2}\right\}
\eqno{(13)}
$$
which are gauge-invariant themselves and for the case $x=0$ they
coincide with the known results (see e.g. [7,8] for SU(2)-group).
$$
\Delta\Omega^{(cor)}/T^4=-\frac{g^3}{4\pi}\sqrt{(\frac{2}{3})^3}
\eqno{(14)}
$$
Unfortunately further only the result (14) has a physical meaning
since all positions of the extremum points (at least for a small g)
are to be the same as in (5).

The corrected points (which should be used for treating the order $g^4$
effective action) are also known [4] and for a small g
these corrections are proportional to the $g^2$-terms
$$
\bar{x}_{1,3}=1\pm\left[1-\frac{g^2}{4{\pi}^2}(1+\frac{1-\xi}{2})\right]
\eqno{(15)}
$$
Being substituted to the lowest orders effective action (3) these points
generate the gauge-dependent $g^4$-corrections
\setcounter{equation}{15}
\begin{eqnarray}
\Omega/T^4&=&\Omega^{(1)}/T^4+\Omega^{(2)}/T^4=2\pi^2B_4(0)
+\frac{3g^2}{2}B_2^2(0)\nonumber\\
&+&\frac{g^4}{48\pi^2}(1+\frac{1-\xi}{2})^2-\frac{g^4}{24\pi^2}
|1+\frac{1-\xi}{2}|-\frac{g^4}{24\pi^2}(1-\xi)|1+
\frac{1-\xi}{2}|\nonumber\\
\end{eqnarray}
and other ones which, however, are beyond the calculational accuracy. So
the nonanalytic $g^3$-terms found above are the leading ones in
the nonperturbative expansion of W(x) and the both expressions (3) and
(13) used jointly are the closed result till the $g^4$-terms are absent.

However all $g^4$-terms (or at least some part of them) should be
calculated exactly to solve the problem of degeneracy as well as
to check with the aid of (15) a gauge-invariance of the order $g^4$
thermodynamical potential. In particular, analysing (16) we consider
that there is a possibility to calculate exactly within the three-loop
graphs all $g^4$-terms which are proportional to $(1-\xi)$-multiplyer
and then to combine these terms with the analogous ones in (16).
Although these terms being put on the extremum points
should be equel zero they are very important to check the necessary
condition of a gauge-invariance for the order $g^4$
thermodynamical potential. Of course it is necessary to find
all other terms which are proportional to the high degrees of
$(1-\xi)$-multiplyer but this task seems to be more complicated and it
can be investigated in the second rate.

The problem of degeneracy should be also solved within $g^4$-terms.
It is doubtless that a trivial vacuum will be splitted but some
degeneracy seems to be kept because this possibility is embeded at once
by a symmetry of the breaking operator to build in accordance with
a structure of the chosen external field. However there are no reasons to
consider this symmetry being proved for the order $g^4$ thermodynamical
potential and a signal about breaking it
will be received if one finds at least a part
of terms which are not proportional to the truncated Bernoulli
polynomials $\tilde {B}_2(z)$ and $\tilde {B}_4(z)$. Here $\tilde
{B}_{2n}=B_{2n}(z)-B_{2n}(0)$ and these functions are periodic under
substitution $|z|\to1-|z|$. Unfortunately these $g^4$-terms
should be calculated directly since any nonperturbative summations with
using the $g^2$-terms keep the found degeneracy at least for a small g.
The $g^3$-terms obtained here display this fact although they are very
important themselves when any applications of the found effective action
are investigated.

References

1. V.M.Belyaev and V.L.Eletsky. Pis'ma Zh. Eksp. Teor. Fis. {\bf 50}
 (1990) 49;

2. V.M.Belyaev. Phys. Lett. {\bf B254} (1991) 153.

3. K.Enqvist and K.Kajantie. Z. Phys. {\bf C47} (1990) 291.

4. O.K.Kalashnikov. Preprint FIAN/TD/06-92 (Moscow, 1992).

5. N.Weiss. Phys. Rev. {\bf D24} (1981) 475;
O.K.Kalashnikov, V.V.Klimov and E.Casado. Phys. Lett. {\bf 114B}
 (1982) 49.  R.Anishetty. J.Phys. {\bf G10} (1984) 439.

6. V.V.Skalozub. Preprint ITP-92-12E (Kiev, 1992).

7. O.K.Kalashnikov and V.V.Klimov, Yad. Fiz. {\bf 33} (1981) 1572
   [Soviet J. Nucl. Phys. {\bf 33} (1981) 847].

8. O.K.Kalashnikov . Fortschr. Phys. {\bf 32} (1984) 525.

\end{document}